\documentclass[12pt]{iopart}

\usepackage{graphics} 
\usepackage{graphicx}
\usepackage{dcolumn}
\usepackage{bm}
\usepackage{subfigure}
\usepackage{multirow}
\usepackage{float} 
\begin{document}

\title{Scaling behaviour of charged hadron $p_{T}$ distributions in $pp$ and $p\bar{p}$ collisions}
\author{W. C. Zhang$^{1}$ and C. B. Yang$^{2}$}
\address{$^{1}$School of Physics and Information Technology, Shaanxi Normal University, Xi'an 710119, People's Republic of China\\
$^{2}$Institute of Particle Physics \& Key Laboratory of Quark and Lepton Physics (MOE), Central China Normal University, Wuhan 430079, People's Republic of China }
\eads{$^{1}$wenchao.zhang@snnu.edu.cn, $^{2}$cbyang@mail.ccnu.edu.cn}
\begin{abstract}

\noindent We report on a scaling behaviour in the transverse momentum ($p_{T}$) distributions for charged hadrons produced in proton-proton ($pp$) collisions with different center of mass energies ($\sqrt{s}$ = 0.9, 2.76 and 7 TeV) at the Compact Muon Solenoid (CMS) detector. This scaling behaviour appears when the $p_{T}$ is replaced by $p_{T}/K$, where $K$ is a parameter and depends on $\sqrt{s}$. A similar scaling behaviour is observed in the $p_{T}$ distributions of charged hadrons produced in proton-antiproton ($p\bar{p}$) collisions with $\sqrt{s}$ = 0.63, 1.8 and 1.96 TeV at the Collider Detector at Fermilab (CDF). The particle production mechanism behind the scaling behaviour in the $pp$ or $p\bar{p}$ collisions could be explained by the model of percolation of strings.

\end{abstract}
\submitto{\jpg}
\maketitle

\section{\label{sec:introduction}Introduction}
One of the main goals in high energy collisions is to investigate the dynamics for particle productions. Several approaches are utilized to search for regularities in the particle production, one of them being the search for a scaling behaviour of some quantities versus suitable variables. 

The scaling behaviour was first introduced in electron-nucleon deep inelastic scatterings (DIS) in which the structure functions depend not on the energy of DIS but on the ratio of the energy to the momentum transfer in the scattering  \cite{DIS, Feynman, KNO}. In recent years, scaling behaviours were observed in nucleus-nucleus collisions. Ref. \cite{pion_spectrum} showed that a scaling behaviour is exhibited in the pion $p_{T}$ spectra with different collision centralities at midrapidity in Au+Au collisions at the Relativistic Heavy Ion Collider (RHIC). This scaling behaviour of pions was also found in noncentral regions in Au+Au and d+Au collisions \cite{non_central_collision}. An analogous scaling behaviour was observed in the proton and anti-proton $p_{T}$ spectra with different collision centralities at midrapidity in Au+Au collisions at RHIC \cite{proton_antiproton_spectra}.

Recently, a universal scaling behaviour was presented in the $p_{T}$ distributions of charged hadrons in $pp$ collisions with $\sqrt{s}$ = 0.9, 2.36 and 7 TeV at CMS \cite{scaling_pp_collision}. This scaling behaviour is seen when the $p_{T}$ distributions are expressed in a suitable variable, $p^{'}_{T}$.  $p^{'}_{T}$ at energy $\sqrt{s^{'}}$ is defined in terms of $p_{T}$ at energy $\sqrt{s}$ and it is written as $p^{'}_{T}=p_{T}(\sqrt{s^{'}}/\sqrt{s})^{\frac{\lambda}{\lambda+2}}$, where $\lambda$ is a parameter and it depends on $p_{T}$ at energy $\sqrt{s}$.  In this paper, we propose another method to search for the scaling behaviour of the charged hadron $p_{T}$ distributions in $pp$ collisions with $\sqrt{s}$ = 0.9, 2.76 and 7 TeV.  This scaling behaviour shows up when the $p_{T}$ distributions are presented in another suitable variable, $z=p_{T}/K$. Here $K$ is a free parameter which only depends on $\sqrt{s}$, rather than $p_{T}$ at energy $\sqrt{s}$.  Similar scaling behaviour is also searched for in the charged hadron $p_{T}$ distributions in $p\bar{p}$ collisions with $\sqrt{s}$ = 0.63, 1.8 and 1.96 TeV at CDF. 

This paper is organized as follows. In \sref{sec:method}, the procedure to search for the scaling behaviour in $pp$ and $p\bar{p}$ collisions is illustrated. \Sref{sec:scaling_behaviour} describes the scaling behaviour of charged hadrons in $pp$ and $p\bar{p}$ collisions with different center of mass energies. \Sref{sec:comparison_between_scaling} shows the comparison between the scaling behaviours presented in the variables $z$ and $p_{T}^{'}$. Finally in \sref{sec:discussions}, the possible particle production mechanism behind the scaling behaviour is discussed.

\section{\label{sec:method}Method to search for the scaling behaviour}

The method to search for the scaling behaviour of charged hadron $p_{T}$ spectra at different energies in $pp$ and $p\bar{p}$ collisions is similar to the one which was described in Refs. \cite{pion_spectrum,non_central_collision, proton_antiproton_spectra}. Here we will describe it briefly. When presented in a suitable variable $z=p_{T}/K$, the scaled $p_{T}$ spectra at differen center of mass energies in $pp$ or $p\bar{p}$ collisions, $\Phi(z)=A\cdot(2\pi p_{T})^{-1}d^{2}N/dp_{T}dy|_{p_{T}=Kz}$, will exhibit a universal scaling behaviour. Here the parameters $A$ and $K$ depend on the energy in the $pp$ or $p\bar{p}$ collisions. While in Refs. \cite{pion_spectrum,non_central_collision, proton_antiproton_spectra}, parameters $A$ and $K$ depend on the centrality of the Au+Au or d+Au collisions. As a convention, $K$ and $A$ are set to be 1 for the highest energy collisions. Obviously, with different choices of $A$ and $K$ for the highest energy collisions, we get different scaling functions $\Phi(z)$. The arbitrariness of $\Phi(z)$ will disappear if the $p_{T}$ spectra at different energies are presented in another variable, $u=z/\langle z \rangle=p_{T}/\langle p_{T} \rangle$. Here $\langle z \rangle=\int^{\infty}_{0}z\Phi(z)zdz\big/\int^{\infty}_{0}\Phi(z)zdz$. The normalized scaling distribution as a function of $u$ is thus defined as $\Psi(u)=\langle z \rangle^{2}\Phi(\langle z \rangle u)\big/\int^{\infty}_{0}\Phi(z)zdz$.
 
\section{\label{sec:scaling_behaviour}Scaling behaviour in $pp$ and $p\bar{p}$ collisions}
The charged hadron $p_{T}$ spectra in $pp$ ($p\bar{p}$) collisions with $\sqrt{s}$ = 0.9, 2.76 and 7 (0.63, 1.8 and 1.96) TeV at CMS (CDF) were published in Refs. \cite{cms_0_9_7,cms_2_76,cdf_0_63_1_8,cdf_1_96}. The $p_{T}$ distributions in CMS (CDF) data cover a range up to 200 (50) GeV/c, which is much larger than the $p_{T}$ coverage of the data presented in Refs. \cite{pion_spectrum,non_central_collision, proton_antiproton_spectra}. Since $A$ and $K$ are both set to be 1 for the highest energy collisions, the scaling function $\Phi(z)$ for the $pp$ collisions is nothing but the charged hadron $p_{T}$ spectrum at $\sqrt{s}$ = 7 TeV. As described in Ref. \cite{pp_collision_tsallis}, the $p_{T}$ spectrum of charged hadrons produced in high energy collisions follows a non-extensive statistical distribution, namely the Tsallis distribution \cite{tsallis_dist_1,tsallis_dist_2,tsallis_dist_3}. Thus the scaling function for the $pp$ collisions can be written as follows: 
\begin{eqnarray}
\Phi(z)=C_{q}\left[1-(1-q)\frac{z}{z_{0}}\right]^{\frac{1}{1-q}},
\label{eq:phi_z_pt_spectrum_pp}
\end{eqnarray}
where $C_{q}$, $q$ and $z_{0}$ are free parameters, and $1-q$ is a measure of the non-extensivity. These parameters are obtained by fitting the $p_{T}$ spectrum at $\sqrt{s}$ = 7 TeV with the Tsallis distribution in Eq. \eref{eq:phi_z_pt_spectrum_pp} using the least $\chi^{2}$s method, and they are tabulated in the second column of \tref{tab:pp_fit_parameters}. The $\chi^2$s divided by the number of degrees of freedom ($dof$), named reduced $\chi^{2}$s, for this fit is 3.7. The scaling parameters $A$ and $K$ for the $pp$ collisions with $\sqrt{s}$ = 0.9 and 2.76 TeV are determined by fitting the scaled Tsallis distribution,
\begin{eqnarray}
\frac{1}{A}\Phi(p_{T}/K)=\frac{C_{q}}{A}\left[1-(1-q)\frac{p_{T}/K}{z_{0}}\right]^{\frac{1}{1-q}},
\label{eq:scaled_phi_z_pt_spectrum_pp}
\end{eqnarray}
to the $p_{T}$ spectra at these two lower collision energies using the least $\chi^{2}$s method. Here only $A$ and $K$ are free parameters, $C_{q}$, $q$ and $z_{0}$ are fixed to the values obtained at $\sqrt{s}$ = 7 TeV. Table \ref{tab:pp_a_k_parameters} presents $A$ and $K$ for the $pp$ collisions with $\sqrt{s}$ = 0.9, 2.76 and 7 TeV. The fits performed on the $p_{T}$ spectra with $\sqrt{s}$ = 0.9 and 2.76 TeV are worse than the fit on the $p_{T}$ spectrum with $\sqrt{s}$ = 7 TeV, which could be seen from the comparison among the reduced $\chi^{2}$s for these three fits.

\begin{figure}[H]
\centering
\includegraphics[scale=0.40]{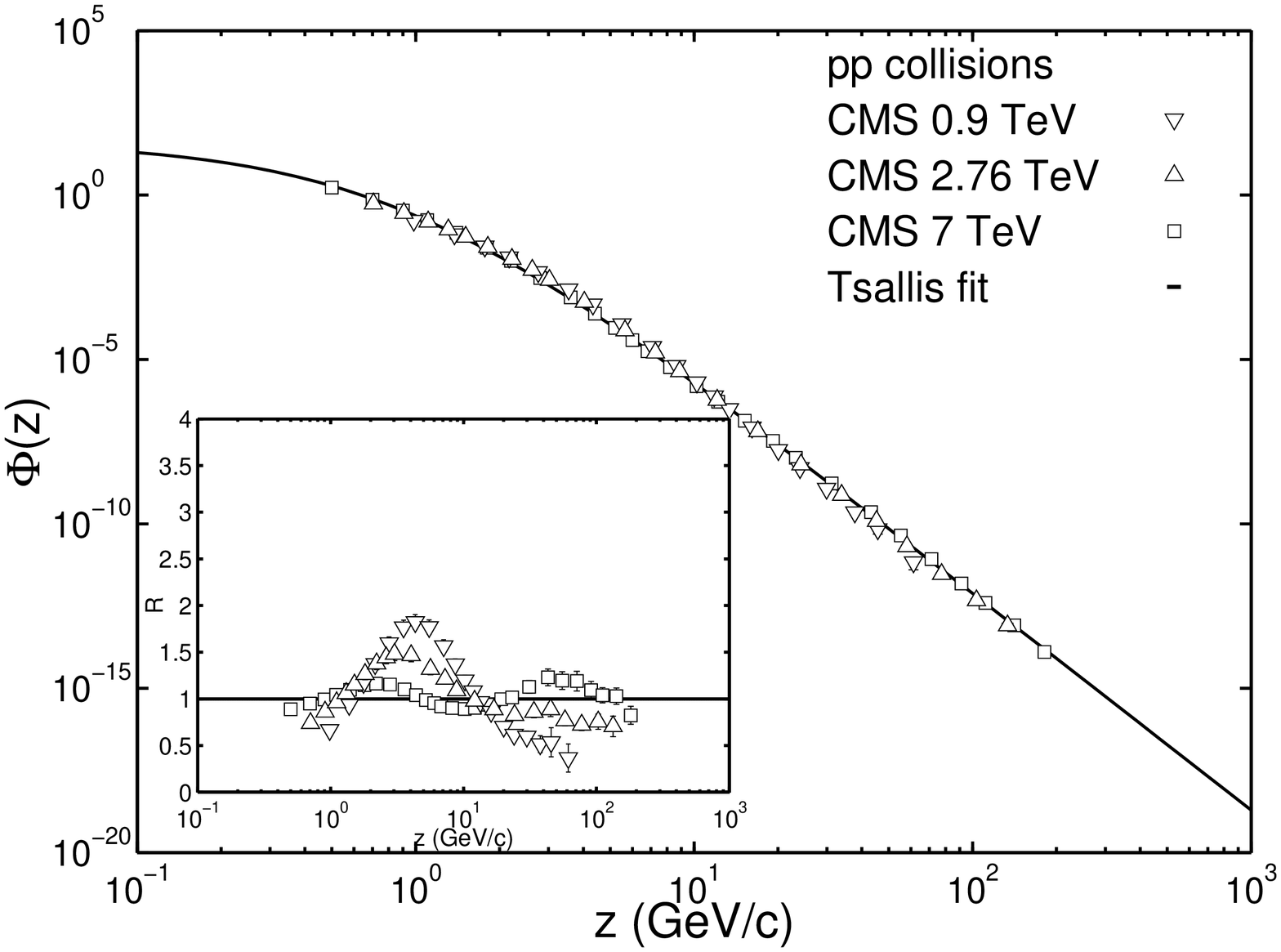} 
\caption{\label{fig:proton_proton_z_plus_ratio_log}Scaling behaviour of the charged hadron $p_{T}$ spectra presented in $z$ for the $pp$ collisions at $\sqrt{s}$ = 0.9, 2.76 and 7 TeV. The solid curve is described by Eq. \eref{eq:phi_z_pt_spectrum_pp} with parameters tabulated in second column of \tref{tab:pp_fit_parameters}. The data points are taken from \cite{cms_0_9_7,cms_2_76}. The inset shows the distribution of the ratio between the experimental data and fitted results.}
\end{figure}

\Fref{fig:proton_proton_z_plus_ratio_log} shows the scaling behaviour of the charged hadron $p_{T}$ spectra presented in $z$ for the $pp$ collisions at $\sqrt{s}$ = 0.9, 2.76 and 7 TeV. In order to see how well the CMS data agree with the fitted curve, we define a ratio, $R=\rm experimental\ data/fitted\ results$. The inset of \fref{fig:proton_proton_z_plus_ratio_log} shows $R$ as a function of $z$ for all data points in the $pp$ collisions with different energies.  Except for a few data points which lie in the moderate $p_{T}$ region with 2 $<z<$ 7 GeV/c, all the data points have $R$ values in the range 0.5-1.5, which implies that the scaling behaviour is true within an accuracy of 50$\%$. Considering the fact that the data in the $pp$ collisions cover about 15 orders of magnitude, the fits performed on the $p_{T}$ spectra with $\sqrt{s}$ = 0.9, 2.76 and 7 TeV are good.

\begin{table}[H]
\caption{\label{tab:pp_fit_parameters} The parameters $C_{q}$, $q$ and $z_{0}$ for the scaling functions $\Phi(z)$ in the $pp$ and $p\bar{p}$ collisions. The uncertainties quoted are statistical. The last line shows the reduced $\chi^{2}$s for the fit performed on the $p_{T}$ spectrum with $\sqrt{s}$ = 7 (1.96) TeV in the $pp$ ($p\bar{p}$) collisions. }
\begin{indented}
\item[] \begin{tabular}{@{}ccc}
\br
\textrm{\ }&
\textrm{$pp$ collisions}&
\textrm{$p\bar{p}$ collisions}\\
\mr
$C_{q}$ & 40.9$\pm$5.1 & 1066.4$\pm$10.2\\
$q$&1.151$\pm$0.001 & 1.123$\pm$0.001\\
$z_{0}$ (GeV/c)& 0.128$\pm$0.004 & 0.154$\pm$0.001\\
$\chi^{2}/dof$& 3.7 & 1.2\\
\br
\end{tabular}
\end{indented}
\end{table}

\begin{table}[H]
\caption{\label{tab:pp_a_k_parameters} Scaling parameters $A$ and $K$ for $pp$ collisions with $\sqrt{s}$ = 0.9, 2.76 and 7 TeV. The uncertainties quoted are statistical. The last column shows the reduced $\chi^2$s for fitting Eq. \eref{eq:scaled_phi_z_pt_spectrum_pp} to the $p_{T}$ spectra at $\sqrt{s}$ = 0.9 and 2.76 TeV. }
\begin{indented}
\item[] \begin{tabular}{@{}cccc}
\br
\textrm{$\sqrt{s}$ (TeV)}&
\textrm{$K$}&
\textrm{$A$}&
$\chi^2/dof$\\
\mr
0.9 & 0.51$\pm$0.05 & 0.15$\pm$0.07&52.3\\
2.76 &0.74$\pm$0.03 & 0.43$\pm$0.11&19.7\\
7 & 1 & 1 &-\\
\br
\end{tabular}
\end{indented}
\end{table}

For the $p\bar{p}$ collisions, the parameters $C_{q}$, $q$ and $z_{0}$ are determined by fitting the Tsallis distribution in Eq. \eref{eq:phi_z_pt_spectrum_pp} to the $p_{T}$ spectrum at $\sqrt{s}$ = 1.96 TeV. These parameters are listed in the third column of \tref{tab:pp_fit_parameters}. The scaling parameters $A$ and $K$ for the $p\bar{p}$ collisions with $\sqrt{s}$ = 0.63, 1.8 and 1.96 TeV are obtained in a similar way as those in the $pp$ collisions and they are presented in \tref{tab:ppbar_a_k_parameters}.

\begin{figure}[h]
\centering
\includegraphics[scale=0.40]{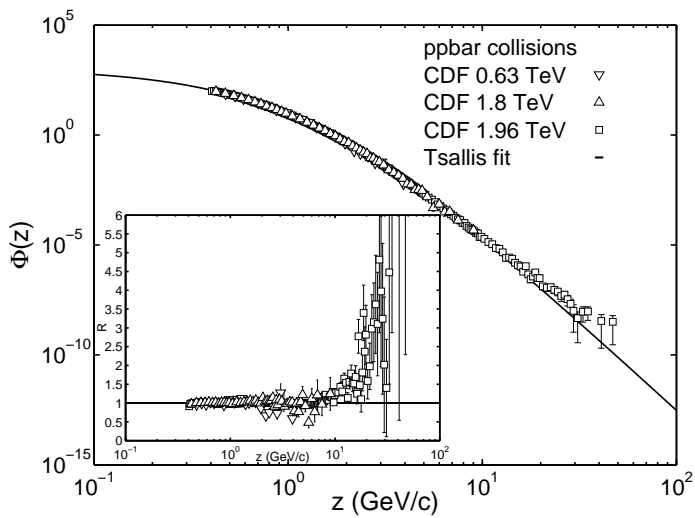} 
\caption{\label{fig:proton_antiproton_z_plus_ratio_log}Scaling behaviour of the charged hadron $p_{T}$ spectra presented in $z$ for the $p\bar{p}$ collisions at $\sqrt{s}$ = 0.63, 1.8 and 1.96 TeV. The solid curve is described by Eq. \eref{eq:phi_z_pt_spectrum_pp} with parameters tabulated in the third column of \tref{tab:pp_fit_parameters}. The data points are taken from \cite{cdf_0_63_1_8,cdf_1_96}. The inset is the distribution of the ratio between the experimental data and fitted results.}
\end{figure}

\Fref{fig:proton_antiproton_z_plus_ratio_log} shows the scaling behaviour of the charged hadron $p_{T}$ spectra presented in $z$ for the $p\bar{p}$ collisions at $\sqrt{s}$ = 0.63, 1.8 and 1.96 TeV. The $R$ distribution as a function of $z$ for the $p\bar{p}$ collisions is presented in the inset of this figure. In the low $p_{T}$ region with $z<10$ GeV/c, the data points and the fitted curve agree within $50\%$. In the high $p_{T}$ region with $z>$ 10 GeV/c, there is a large deviation between the data points and the fitted curve.

\begin{table}[H]
\caption{\label{tab:ppbar_a_k_parameters} Scaling parameters $A$ and $K$ for $p\bar{p}$ collisions with $\sqrt{s}$ = 0.63, 1.8 and 1.96 TeV. The uncertainties quoted are statistical. The last column shows the reduced $\chi^2$s for the fits performed on the  $p_{T}$ spectra at $\sqrt{s}$ = 0.63 and 1.8 TeV.}
\begin{indented}
\item[]\begin{tabular}{@{}cccc}
\br
\textrm{$\sqrt{s}$ (TeV)}&
\textrm{$K$}&
\textrm{$A$}&
$\chi^{2}/dof$\\
\mr
0.63 & 0.89$\pm$0.02 & 2.85$\pm$0.18&2.0\\
1.8 &0.999$\pm$0.004 & 2.28$\pm$0.03&2.7\\
1.96& 1 & 1& -\\
\br
\end{tabular}
\end{indented}
\end{table}

\begin{figure}[h]
\centering
\mbox
{
\subfigure[]{\label{fig:tsallis_proton_proton_u_log_least_squares}{\includegraphics[scale=0.35]{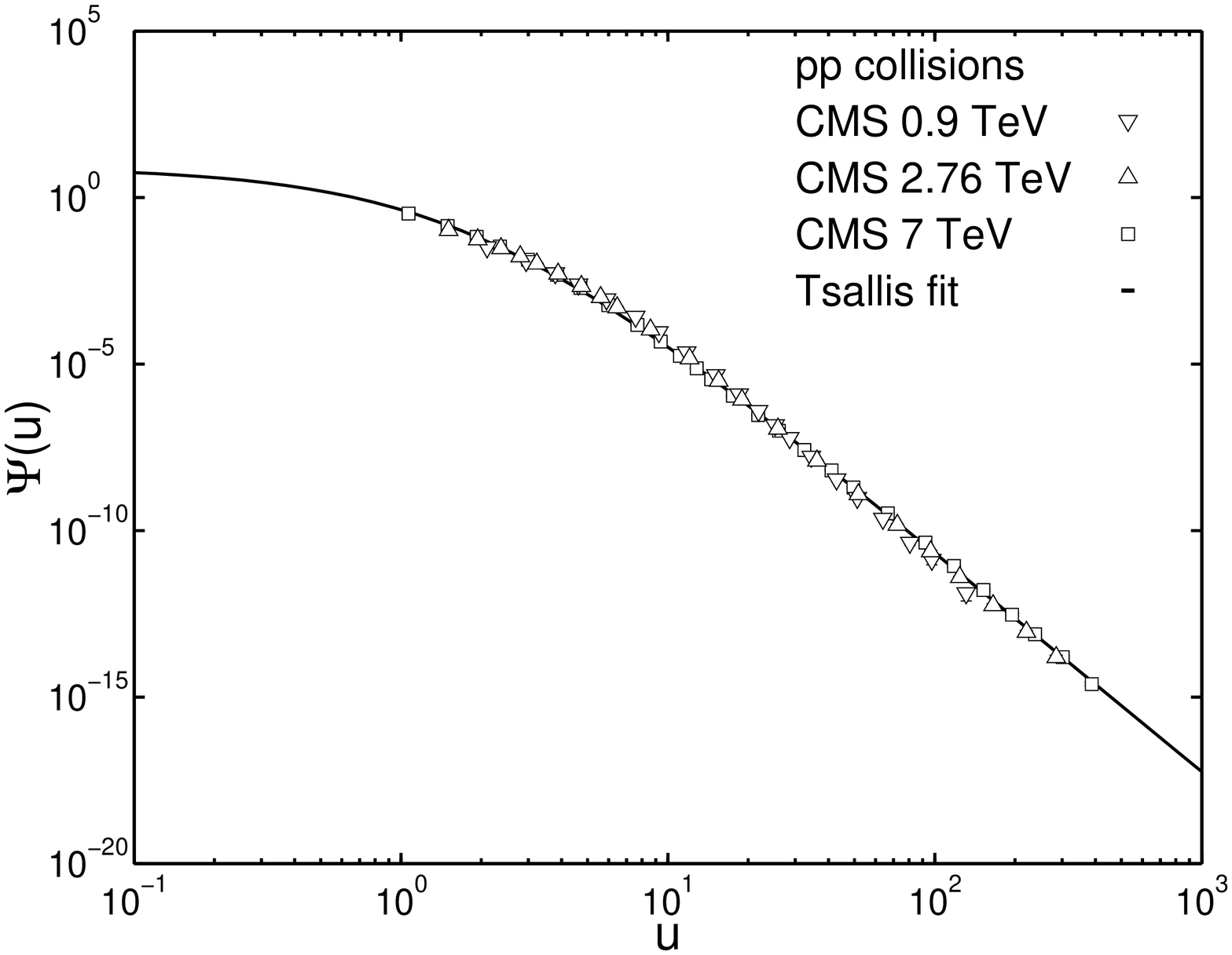}}
}
\subfigure[]{\label{fig:tsallis_proton_antiproton_u_log_least_squares}{\includegraphics[scale=0.345]{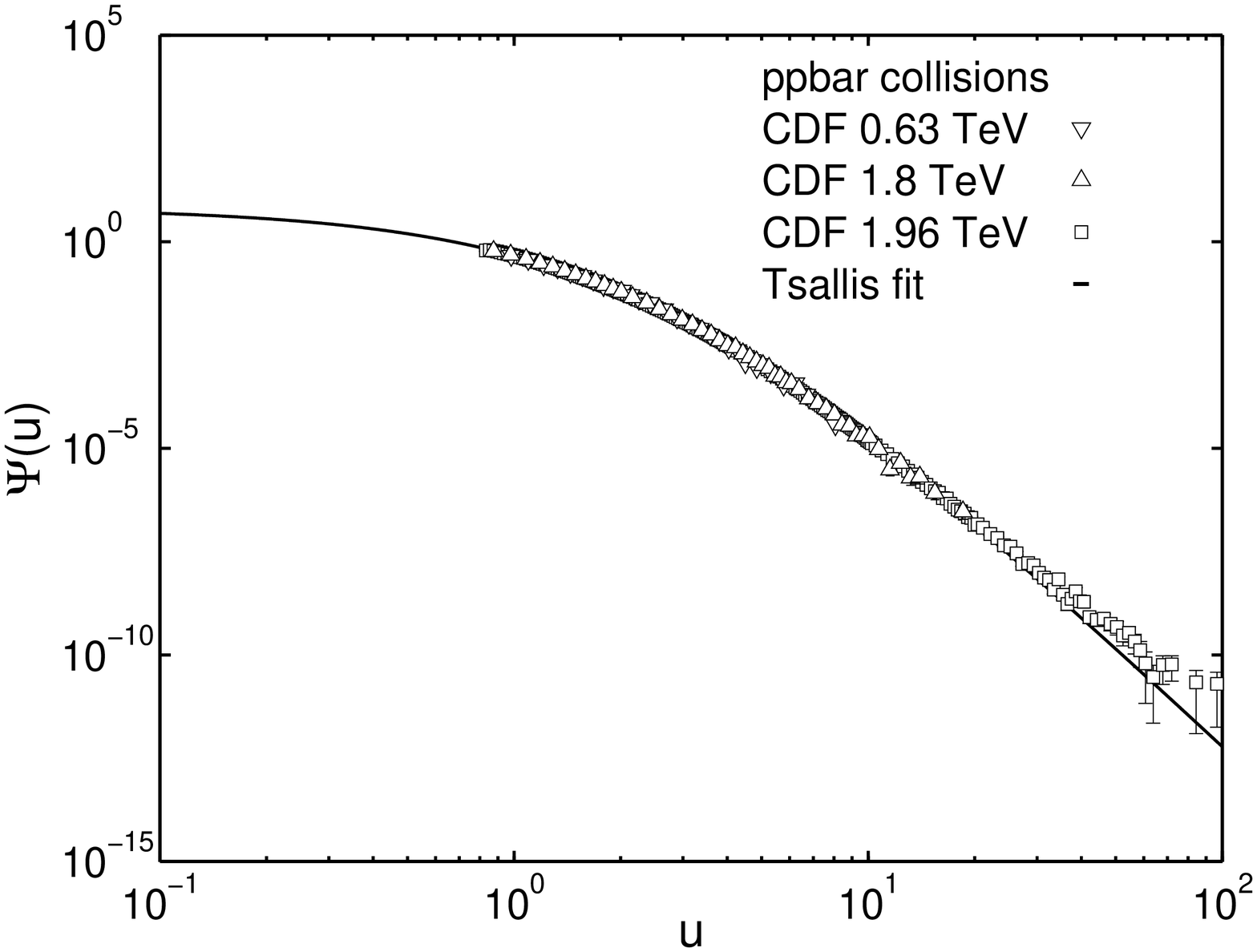}}
}
}
\caption{\label{fig:tsallis_proton_(anti)proton_u_log_dist_least_squares} (a) ((b)), normalized scaling distribution as a function of variable $u$ for the charged hadrons produced in the $pp$ ($p\bar{p}$) collisions. The solid curves are described by Eq. \eref{eq:phi_u_pt_spectrum_pp}. The CMS (CDF) data points are taken from \cite{cms_0_9_7,cms_2_76} (\cite{cdf_0_63_1_8,cdf_1_96}).}
\end{figure}

So far we have seen that the charged hadron $p_{T}$ distributions in the $pp$ and $p\bar{p}$ collisions indeed exhibit a scaling behaviour. However, the scaling function in Eq. \eref{eq:phi_z_pt_spectrum_pp} relies on the choice of parameters $A$ and $K$ for the highest energy collisions. In order to eliminate this dependence on the choice, the scaling variable $z$ is replaced by $u=z/\langle z\rangle$. In the $pp$ ($p\bar{p}$) collisions, $\langle z \rangle$ for the charged hadrons is determined as 0.47$\pm$0.02 (0.486$\pm$0.002) GeV/c with the definite integral of $z$ over the interval [0, 200] ([0,50]) GeV/c, which roughly corresponds to the $p_{T}$ range measured by CMS (CDF). Plugging $\langle z \rangle$ as well as $\Phi(z)$ into $\Psi(u)$ defined in \sref{sec:method}, one can easily get the normalized scaling functions $\Psi(u)$ (see \fref{fig:tsallis_proton_(anti)proton_u_log_dist_least_squares}) for the $pp$ and $p\bar{p}$ collisions as follows: 
\begin{eqnarray}
\Psi(u)=C^{'}_{q}\left[1-(1-q^{'})\frac{u}{u_{0}}\right]^{\frac{1}{1-q^{'}}}.
\label{eq:phi_u_pt_spectrum_pp}
\end{eqnarray}
Here the parameters $C^{'}_{q}$, $q^{'}$ and $u_{0}$ in the $pp$ ($p\bar{p}$) collisions are determined from $C_{q}$, $q$ and $z_{0}$ at $\sqrt{s}$ = 7 (1.96) TeV using the following expressions: $C^{'}_{q}=\langle z \rangle^{2}C_{q}/\int^{\infty}_{0}\Phi(z)zdz$, $q^{'}=q$ and $u_{0}=z_{0}/\langle z \rangle$. Table \ref{tab:pp_fit_parameters_psi_u} tabulates $C^{'}_{q}$, $q^{'}$ and $u_{0}$ for $\Psi(u)$ in the $pp$ and $p\bar{p}$ collisions.

\begin{table}[H]
\caption{\label{tab:pp_fit_parameters_psi_u} The parameters $C^{'}_{q}$, $q^{'}$ and $u_{0}$ for the normalized scaling functions $\Psi(u)$ in the $pp$ and $p\bar{p}$ collisions. The uncertainties quoted are due to the errors of parameters $C_{q}$, $q$ and $z_{0}$ at $\sqrt{s}$ = 7 and 1.96 TeV.}
\begin{indented}
\item[] \begin{tabular}{@{}ccc}
\br
\textrm{\ }&
\textrm{$pp$ collisions}&
\textrm{$p\bar{p}$ collisions}\\
\mr
$C^{'}_{q}$ & 7.9$\pm$1.2 & 6.63$\pm$0.07\\
$q^{'}$&1.151$\pm$0.001 & 1.123$\pm$0.001\\
$u_{0}$ & 0.27$\pm$0.01& 0.316$\pm$0.001\\
\br
\end{tabular}
\end{indented}
\end{table}

The scaling behaviour of the $p_{T}$ spectra for charged hadrons produced in the $pp$ and $p\bar{p}$ collisions could be validated experimentally in the following way. With the normalized scaling functions in Eq. \eref{eq:phi_u_pt_spectrum_pp}, we can calculate the ratio between the moments of the momentum distributions,
\begin{eqnarray}
\frac{\langle p_{T}^{n}\rangle}{\langle p_{T}\rangle^{n}}=\int^{\infty}_{0}u^{n}\Psi(u)udu,
\label{eq:ratio_p_t_moment}
\end{eqnarray}
where $n=2,3,4,\cdots$. The integration interval for $\Psi_{pp}(u)$ ($\Psi_{p\bar{p}}(u)$), which corresponds to the range of $p_{T}/\langle p_{T}\rangle$ measured by CMS (CDF), is from 0 to 400 (100). Table \ref{tab:ratio_p_t_moment} tabulates $\langle p_{T}^{n}\rangle/\langle p_{T}\rangle^{n}$ with $n=2,3,4$ for the charged hadrons produced in the $pp$ and $p\bar{p}$ collisions. Judging from Eq. \eref{eq:ratio_p_t_moment}, $\langle p_{T}^{n}\rangle/\langle p_{T}\rangle^{n}$ does only depend on the form of the normalized scaling function $\Psi(u)$ in the $pp$ or $p\bar{p}$ collisions. If the scaling behaviour of the charged hadron $p_{T}$ spectra exists, then the ratio between the moments of momentum should be a constant for collisions with different energies. As an example, the ratios between the moments of momentum with $n=2$ are calculated using a combination of the measured data points at $\sqrt{s}$ = 0.9, 2.76 and 7 TeV and the low and high $p_{T}$ contributions determined from the fits of the Tsallis distribution in Eq. \eref{eq:phi_z_pt_spectrum_pp} to the $p_{T}$ spectra at these energies. After a detailed calculation, the values of $\langle p_{T}^{2}\rangle/\langle p_{T}\rangle^{2}$ for the $pp$ collisions with $\sqrt{s}$ = 0.9, 2.76 and 7 TeV are 1.9$\pm$0.1, 2.0$\pm$0.3 and 2.1$\pm$0.2. Here the uncertainties quoted are due to the errors on the data points as well as the contributions in the high and low $p_{T}$ regions. These values and the value calculated with the integral of the normalized scaling function $\Psi(u)$ in the $pp$ collisions, 2.1$\pm$0.3, are deemed to be consistent within uncertainties. This agreement confirms that the scaling behaviour of the charged hadron $p_{T}$ spectra in the $pp$ collisions at the CMS energies is true. We can also test the scaling behaviour of the charged hadron $p_{T}$ spectra in the $p\bar{p}$ collisions at the CDF energies in a similar way.

\begin{table}[h]%
\caption{\label{tab:ratio_p_t_moment}$\langle p_{T}^{n}\rangle/\langle p_{T}\rangle^{n}$ calculated with Eq. \eref{eq:ratio_p_t_moment} for the charged hadrons produced in the $pp$ and $p\bar{p}$ collisions. The uncertainties quoted are due to the errors of parameters $C^{'}_{q}$, $q^{'}$ and $u_{0}$. }
\begin{indented}
\item[]\begin{tabular}{@{}ccc}
\br
\textrm{$n$}&
\textrm{$pp$ collisions}&
\textrm{$p\bar{p}$ collisions}\\
\mr
2& 2.1$\pm$0.3 & 1.86$\pm$0.03 \\
3 & 9.3$\pm$1.4 & 6.1$\pm$0.1\\
4 &121.4$\pm$20.7 & 36.4$\pm$1.0\\
\br
\end{tabular}
\end{indented}
\end{table}

\begin{figure}[H]
\centering
\includegraphics[scale=0.40]{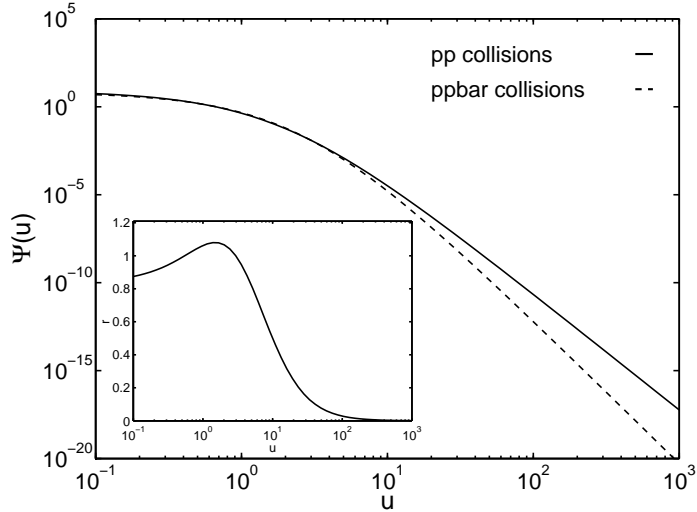} 
\caption{\label{fig:comp_u_pp_ppbar_tsallis_diff} Comparison between the normalized scaling functions $\Psi(u)$ in the $pp$ and $p\bar{p}$ collisions. The inset is the distribution of $r$ as a function of $u$.}
\end{figure}

Now one would like to ask for the difference between the scaling functions in the $pp$ and $p\bar{p}$ collisions. Figure \ref{fig:comp_u_pp_ppbar_tsallis_diff} shows the comparison between the two normalized scaling functions $\Psi(u)$ for the $pp$ and $p\bar{p}$ collisions. When $u$ is small, the discrepancy between these two scaling functions is not so obvious in logarithm scale. In order to present it clearly, we define a variable $r=\Psi_{p\bar{p}}(u)/\Psi_{pp}(u)$ and show its distribution as a function of $u$ in the inset of Fig. \ref{fig:comp_u_pp_ppbar_tsallis_diff}. When $u$ is large, $r$ decreases with the increase of $u$ monotonically. This is confirmed by the comparison between the values of $\langle p_{T}^{n}\rangle/\langle p_{T}\rangle^{n}$ in the $pp$ and $p\bar{p}$ collisions (see \tref{tab:ratio_p_t_moment}). Because of small difference between these two normalized scaling functions at small $u$, the value of $\langle p_{T}^{n}\rangle/\langle p_{T}\rangle^{n}$ in the $pp$ collisions is close to that value in the $p\bar{p}$ collisions for small $n$. However, for large $n$, the value of $\langle p_{T}^{n}\rangle/\langle p_{T}\rangle^{n}$ in the $pp$ collisions differs significantly from the value in the $p\bar{p}$ collisions due to the big difference between these two normalized scaling functions  at large $u$.
 
\section{\label{sec:comparison_between_scaling} Comparison between the scaling behaviours presented in $z$ and $p_{T}^{'}$}
\begin{figure}[h]
\centering
\includegraphics[scale=0.40]{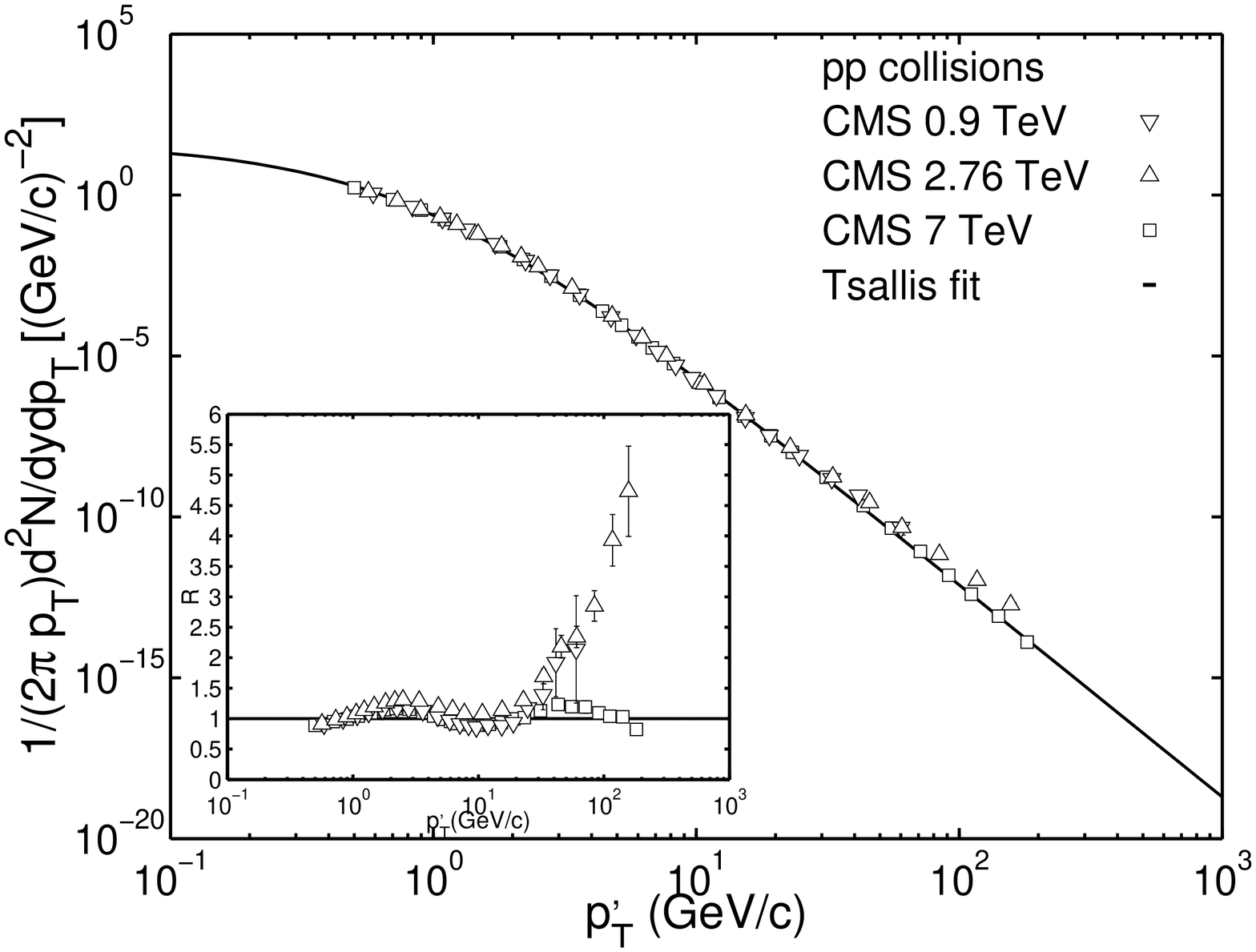} 
\caption{\label{fig:pt_prime_p_p_scaling}Scaling behaviour of the charged hadron $p_{T}$ spectra presented in $p_{T}^{'}$ in the $pp$ collisions with $\sqrt{s}$ = 0.9, 2.76 and 7 TeV. The solid curve is described by Eq. \eref{eq:phi_z_pt_spectrum_pp} with parameters in the second column of \tref{tab:pp_fit_parameters}. The data points are taken from \cite{cms_0_9_7,cms_2_76}. The inset shows the distribution of the ratio between the experimental data and fitted results.}
\end{figure}

As described in Ref. \cite{scaling_pp_collision}, there is a scaling behaviour when the charged hadron $p_{T}$ spectra in the $pp$ collisions at the CMS energies are presented in the variable $p_{T}^{'}$. The $p_{T}$ spectrum at $\sqrt{s}$ is connected with the $p_{T}^{'}$ spectrum at $\sqrt{s^{'}}$ via $p^{'}_{T}=p_{T}(\sqrt{s^{'}}/\sqrt{s})^{\frac{\lambda}{\lambda+2}}$, where $\lambda=0.13+0.1(4p_{T}^{2}/10)^{0.35}$. \Fref{fig:pt_prime_p_p_scaling} shows the $p_{T}$ distributions presented in $p_{T}^{'}$ in the $pp$ collisions with $\sqrt{s}$ = 0.9, 2.76 and 7 TeV. In this figure, the $p_{T}$ spectra at $\sqrt{s}$ = 2.76 and 0.9 TeV have been rescaled to the $p_{T}^{'}$ spectrum at $\sqrt{s^{'}}$ = 7 TeV. The latter spectrum is described  by the Tsallis fit in Eq. \eref{eq:phi_z_pt_spectrum_pp} with parameters in the second column of \tref{tab:pp_fit_parameters}. The inset of \fref{fig:pt_prime_p_p_scaling} shows the distribution of the ratio between the experimental data and fitted results. When $p_{T}^{'}$ is smaller than 20 GeV/c, the scaling behaviour of the charged hadron $p_{T}$ spectra presented in $p_{T}^{'}$ is true within an accuracy of 50$\%$. This accuracy is comparable to the accuracy of the scaling behaviour of the charged hadron $p_{T}$ spectra presented in $z$. However, when $p_{T}^{'}$ is greater than 20 GeV/c, the $p_{T}$ spectra at $\sqrt{s}$ = 2.76 and 0.9 TeV start to deviate significantly from the $p_{T}$ spectrum at $\sqrt{s}$ = 7 TeV.

\section{\label{sec:discussions}Discussions}
We have shown that there is a scaling behaviour in the $p_{T}$ distributions of charged hadrons produced in the $pp$ ($p\bar{p}$) collisions at the CMS (CDF) energies. This scaling behaviour is observed when a linear transformation is applied on $p_{T}$. The Tsallis distribution can be applied to describe the universal shape of the spectra behind this scaling behaviour.  In order to understand the particle production mechanism behind this scaling behaviour, the model of percolation of strings is utilized \cite{percolaton_string}. In this model, color strings are stretched between the two colliding hadrons in the $pp$ or $p\bar{p}$ collisions. These strings will split into new strings with the emission of $q\bar{q}$ pairs. Observed hadrons are formed through this quark pair emission. The transverse area of a color string  is $S_{1}=\pi r_{0}^{2}$, where $r_{0}=0.2$ fm. If there are $n$ strings, they may overlap with each other and thus form a cluster with  a transverse area of $S_{n}$. The $p_{T}$ distribution at energy $\sqrt{s^{'}}$ could be related to the $p_{T}$ distribution at energy $\sqrt{s}$ by a linear transformation on $p_{T}$ at energy $\sqrt{s^{'}}$: $p_{T} \rightarrow p_{T}/((nS_{1}/S_{n})_{\sqrt{s^{'}}}/(nS_{1}/S_{n})_{\sqrt{s}})^{1/4}$. Here $nS_{1}/S_{n}$ gives the degree of string overlap. If strings just get in touch with each other, then $S_{n}=nS_{1}$ and $nS_{1}/S_{n}=1$. If strings maximumly overlap with each other, then $S_{n}=S_{1}$ and $nS_{1}/S_{n}=n$ with $n>1$. Comparing the $p_{T}$ transformation in this model with the one used in the way to search for the scaling behaviour, $p_{T}\rightarrow p_{T}/K$, we know that $K$ gives the ratio between the degrees of string overlap for the collisions at $\sqrt{s^{'}}$ and $\sqrt{s}$. For the $pp$ ($p\bar{p}$) collisions at CMS (CDF), $\sqrt{s}$ is set to be 7 (1.96) TeV and $\sqrt{s^{'}}$ is set to be 0.9, 2.76 and 7 (0.63, 1.8 and 1.96) TeV. As described in Ref. \cite{percolaton_string}, the degree of string overlap, $nS_{1}/S_{n}$, grows with the increase of the energy. Thus $K$ should also grow with the increase of the energy. That's indeed what we observed in tables \ref{tab:pp_a_k_parameters} and \ref{tab:ppbar_a_k_parameters}. We found that in the $pp$ ($p\bar{p}$) collisions the degrees of string overlap at $\sqrt{s^{'}}$ = 0.9 and 2.76 (0.63 and 1.8) TeV are (6.8$\pm$2.7)$\%$ and (30.0$\pm$4.9)$\%$ ($(62.7\pm5.6)\%$ and $(99.6\pm1.6)\%$) of the degree of string overlap at $\sqrt{s}$ = 7 (1.96) TeV. As a summary, the scaling behaviour we observed in the $p_{T}$ distributions of charged hadrons produced at different colliding energies can be phenomenologically understood within the string percolation model.

\ack
This work was supported by Shaanxi Normal University. This work was also supported in part by the National Natural Science Foundation of China under Grant Nos. 11075061 and 11221504, by the Ministry of Education of China under Grant No.306022, and by the Programme of Introducing Talents of Discipline to Universities under Grant No. B08033.


\end{document}